\documentclass[preprint,showpacs,preprintnumbers,amsmath,amssymb]{revtex4}

\usepackage{graphicx}
\usepackage{dcolumn}
\usepackage{bm}
\usepackage[]{amsmath,amssymb,epsfig,color}
\newcommand{\be}{\begin{equation}}
\newcommand{\ee}{\end{equation}}
\begin{document}


\title{Entanglement of a mixed state of two qubit system}

\author{L. Chotorlishvili}

%

 \affiliation{Institute for Physik, Universitat Augsburg, 86135 Augsburg,Germany}

\begin{abstract}
Interaction with environment may lead to the transition of quantum system
from pure state to the mixed one. In this case, the problem of definition
of entanglement may arise. In particular, quantitative measure
of entanglement concurrence is introduced differently for pure and mixed
states.
    To solve this problem, alternative definition of concurrence
of mixed states is introduced in  present work. In particular,
concurrence of mixed states is determined as an average of pure
states concurrences. Averaging is carried out over the quantum
ensemble, corresponding to all possible realization of random
variable.
\end{abstract}
\pacs{0367.Mn.89.70.+c}
 \maketitle
 When studying complex quantum systems, characteristics, related
 to the correlation of quantum states, often manifest themselves. This
 leads to necessity of introduction and consideration of
 entangled states. The first attempts to analyze characteristics
 of entangled states were made by Einstein  \cite{Einstein} and
 Schr\"{o}dinger \cite{Schrodinger} in the 30-th  of the 20-
 century. In spite of traditional fundamental significance of
 entangled state, the present splash of interest  is
 motivated by possible application to the new physical problems. In recent years
 entangled states have become important concept of new applied
 disciplines such as quantum cryptography \cite{Gisin}, quantum
 theory of information \cite{Childs}, and physics of quantum
 computation \cite{Amico}. It is well known, that entanglement is
 a specific quantum form of correlation, which, however, possesses
 a number of essential differences from classical correlations
 \cite{Carvalho, Guo, Bennett, Schlegel, Su, Garraway, Wang, Tolnikov,
 Mintert}. That is why, it describes exclusively a state of
 quantum mechanical system and defuse classical description.
    Entanglement of spin systems is of special interest
 \cite{Schenk, Liang}. This kind of problems arise in the field of
 Cavity Quantum Electrodynamics (CQED) \cite{Aoki,Mabuchi, Hood}
 for example. This field, in its turn, is one of the promising
 trends for quantum computing  \cite{Raimond,
 Turchette}.
    As is known \cite{Carvalho}, quantum correlations are
 connected with coherent superposition states of compound quantum
 systems. That is why, the first simplest definitions of entangled
 states were introduced for pure quantum mechanical states
 \cite{Carvalho}. In particular, it turned out, that quantum
 correlation could be described with the aid of concurrence.
 Concurrence was originally introduced as an auxiliary
 quantity, used to calculate the entanglement of formation of
 $2\times2$ systems. However, concurrence can also be considered
 as an independent entanglement measure
 \cite{Chotorlishvili,Skrinnikov}. The original definition of
 concurrence  for bipartite two-level systems is given in terms
 of a special basis \cite{Carvalho}:
 \be
 |e_{1}\rangle=|\Phi^{+}\rangle,~~~~~~|e_{2}\rangle=i|\Phi^{-}\rangle,~~~~~~~~
 |e_{3}\rangle=i|\Psi^{+}\rangle,~~~~~~~~
 |e_{4}\rangle=|\Psi^{-}\rangle,
 \label{eq:bas}
 \ee
 where
 $|\Phi^{\pm}\rangle=\big(|00\rangle\pm|11\rangle\big)/\sqrt{2}$
 and
 $|\Psi^{\pm}\rangle=\big(|01\rangle\pm|10\rangle\big)/\sqrt{2}$ are
 the Bell states \cite{Amico}.
    Using this particular basis, the concurrence of a pure state
 $|\Psi\rangle$ is defined as:
 \be
 C\big(\Psi\big)=\big|\sum\limits_{i}\langle
 e_{i}|\Psi\rangle^{2}\big|.
 \label{eq:state}
 \ee
    The value of $C(\Psi)$ determines the existence of quantum
 correlation in the system . in particular the condition
 $C(\Psi)=0$ corresponds to the absence of correlation. The
 maximum value of concurrence is reached for Bell states and is
 equal to $C(e_{j})=\big|\sum\limits_{i}\langle
 e_{i}|e_{j}\rangle^{2}\big|=1$.
    But pure entangled states are only the simplest case of
 quantum  correlated systems. Generalization of introduced
 concepts and characteristics, for the case of mixed states, is
 non trivial and often equivocal.
    Formally, entanglement of mixed state is determined as minimal
 average value of entanglement of ensemble of pure states, that
 realize the given mixed state.
    The concurrence of mixed states is given by \cite{Wootters}:
 \be \label{eq:mixed}
 C_{mixed}(\rho)=\inf\limits_{\{P_{i},\Psi_{i}\}}\sum\limits _{i}P_{i}C(\Psi_{i}),
 \ee
 where
 \be
\rho=\sum\limits_{i}P_{i}|\Psi_{i}\rangle\langle\Psi_{i}|,
~~~~~~P_{i}>0. \label{eq:rho}
 \ee
 Formula (\ref{eq:mixed}) makes it possible to compute concurrence
 of mixed states for arbitrary states of two qubits system. In
 particular, as was shown in \cite{Wootters,Hill}
 \be
 C_{mixed}(\rho)=\max\{0,-\lambda_{1},-\lambda_{2},-\lambda_{3},-\lambda_{4}\}
 \ee
 where $\lambda_{i}$ are the eigenvalues, in decreasing order, of
 the Hermitian matrix
 $R\equiv\sqrt{\sqrt{\rho}\widetilde{\rho}\sqrt{\rho}}$ and $\widetilde{\rho}=(\sigma_{y}\otimes\sigma_{y})\rho^{*}(\sigma_{y}\otimes\sigma_{y})$.
 But as we have noted definition of concept of entanglement for
 mixed state is not trivial. When studying non-stationary dynamics
 of chaotic quantum mechanical systems, subjected to the action of  environment,
 the following scenario may be realized:
     At the initial moment of time $t=0$, the system being  in the
 pure quantum mechanical state, during time evolution performs  irreversible transition to mixed state
 \cite{Skrinnikov, Toklikishvili}. In this case, is not clear
 which definition (\ref{eq:state}) or (\ref{eq:mixed}) needs to
 be applied. Since, at the different moments of time, system is in
 different states.
    In this work we shall try to introduce alternative, more
 adequate for non-stationary chaotic systems, definition of
 concurrence.
    With that end in view we shall consider generalized Jaynes-Cummings
 model, introduced in the work \cite{Chotorlishvili}.
 This model describes a system of two interacting qubits, under the action of cavity field.
 Mixed state is formed due to the averaging over quantum ensemble
 corresponding to all possible realizations of
 stochastic parameter. As it will be shown below, random
 trajectory of a system plays a role of this parameter.
    It should also be noted, that the values of mixed state
 parameters $P_{i}$ in (\ref{eq:rho}) are determined exactly in our
 case. That is why we can not consider them as free parameters, with respect to which
 the expression (\ref{eq:mixed})  should be minimized (As requires common definition of
concurrence of mixed states  \cite{Carvalho} ) . This is one
 more confirmation of necessity of introduction of alternative
 definition of concurrence for non-stationary chaotic systems.
    As we have noted, the system we are interested in, presents itself the generalized
 Jaynes-Cummings model for the case of two interacting
 two-level atoms \cite{Chotorlishvili}.
    The Hamiltonian of the problem under study has the following
 form:
 \begin{eqnarray}\label{eq:Ham}
 && \hat{H}=\frac{\hat{p}^{2}}{2m}+\hat{H}_{s}+\hat{H}_{SB}+\hat{H}_{B},\nonumber
 \\
&& \hat{H}_{S}=\omega_{0}(\hat{S}_{1}^{z}+\hat{S}_{2}^{z})+\Omega(\hat{S}_{1}^{+}\hat{S}_{2}^{-}+\hat{S}_{1}^{-}\hat{S}_{2}^{+}),\\
 && \hat{H}_{SB}=-g(x)\big((\hat{S}_{1}^{+}+\hat{S}_{2}^{+})\hat{b}+(\hat{S}_{1}^{-}+\hat{S}_{2}^{-})\hat{b}^{+}\big),
 \nonumber
 \end{eqnarray}
 where $\hbar=1$, $\omega_{0}$ is the Zeeman frequency of the
 spins located in the field inside the resonator, $\Omega$ is a
 constant of interaction between spins in frequency units,
 $\hat{S}^{z}_{1,2}$, $\hat{S}_{1,2}^{\pm}$ are spin operators of
 two-level atoms. The last term in (\ref{eq:Ham}) is the
 Hamiltonian of the field:
 $\hat{H}_{B}=\omega_{f}\hat{b}^{+}\hat{b}$, $\hat{b}^{+}$ and
 $\hat{b}$ are respectively the photon creation operator and the
 annihilation operator.
    As was shown in \cite{Chotorlishvili}, due to nonlinear
 dependence of the interaction constant  on the system's coordinate
 $g(x)=g_{0}\cos(kx)$, dynamics has very complex and
 chaotic character. Time dependence $x(t)$ itself may be
 considered as random Wiener process. This makes a definite
 influence upon quantum mechanical time evolution of the system \cite{Toklikishvili, Chotorlishvili}.
    As a result, the problem is reduced to the stochastic Schr\"{o}dinger
equation, having the following form in interaction
representation: \be
\label{eq:schrodinger}i\frac{d|\hat{\Psi}(t)\rangle}{dt}
=\hat{V}|\hat{\Psi}(t)\rangle, \ee where $\hat{V}$ is a potential,
stochasticity of which is provided by chaotic time
dependence of the coordinate of the system  $x(t)$ :  \be
\label{eq:potential}
\hat{V}=\Omega(\hat{S}_{1}^{+}\hat{S}_{2}^{-}+\hat{S}_{1}^{-}\hat{S}_{2}^{+})+\omega_{f}\hat{b}^{+}\hat{b}-
g_{0}\cos(k_{f}x)\big((\hat{S}_{1}^{+}+\hat{S}_{2}^{+})\hat{b}+(\hat{S}_{1}^{-}+\hat{S}_{2}^{-})\hat{b}^{+}\big)
\ee
    Assume, that at the time zero $(t=0),$ the wave function
represents itself the direct product of system and fields wave
functions: \be \label{eq:wave}
|\Psi(t=0)\rangle=|\Psi_{system}\rangle\otimes|\Psi_{field}\rangle\ee .
Here
\begin{eqnarray} \label{eq:system}
&& |\Psi_{system}\rangle =
C_{00}|00\rangle+C_{01}|01\rangle+C_{10}|10\rangle+C_{11}|11\rangle ,
\nonumber \\
&& |\Psi_{field}\rangle=\sum\limits_{n}W_{n}|n\rangle .
\end{eqnarray}
Due to interaction (\ref{eq:potential}), only following
transitions between states are possible:
\begin{eqnarray} \label{eq:transition}
&&
|0,0,n+1\rangle\leftrightarrow|0,1,n\rangle,~~~~~~|0,0,n+1\rangle\leftrightarrow|1,0,n\rangle\
\nonumber \\
&&
|0,1,n\rangle\leftrightarrow|1,1,n-1\rangle,~~~~~~|1,0,n\rangle\leftrightarrow|1,1,n-1\rangle\
\\
&&|1,0,n\rangle\leftrightarrow|0,1,n\rangle. \nonumber
\end{eqnarray}
    On the basis of equations (\ref{eq:transition}), we shall
search for the solution of equation (\ref{eq:schrodinger}) in the
following form: \be \label{eq:sol}
|\Psi(t)\rangle=\sum\limits_{n}C_{0,0,n+1}|0,0,n+1\rangle+\sum\limits_{n}C_{0,0,n}|0,1,n\rangle+
\sum\limits_{n}C_{1,0,n}|1,0,n\rangle+\sum\limits_{n}C_{1,1,n-1}|1,1,n-1\rangle.\ee
Substituting Eq.(\ref{eq:sol}) into Eq.(\ref{eq:schrodinger})
and taking into account Eq.(\ref{eq:potential}), one can deduce
system of differential equations, for time dependent coefficients of resolution of wave function.
Corresponding calculations takes up a lot of space and for brevity
we don't bring them here \cite{Chotorlishvili}.
    But here we note that, time dependence of the
coefficients of wave function in (\ref{eq:sol}) are determined by
the functional: \be \label{eq:Q}
Q[\omega(t)]=e^{i\int\limits_{0}^{t}\omega(t')dt'}\ee  where \be
\label{omega} \omega(t)=\sqrt{2(2n+1)}g_{0}\cos(k_{f}x(t)). \ee On
the other hand, as was shown in \cite{Toklikishvili}, because of
the chaotic dynamics, $x(t)$ may be treated as a classical random
process. In this case, for the determination of the systems state
it is necessary to average the functional in (\ref{eq:Q}) by all
realizations of the stochastic variable $x(t)$. This leads to the
transition of the system from pure state to the mixed one. Since
non-diagonal matrix elements of density matrix
\cite{Toklikishvili, Chotorlishvili}: \be \label{eq:density}
\rho_{ijkl}=\rho_{system}\otimes\rho_{field}=C_{a_{i}n_{j}}\cdot
C^{*}_{a_{k}n_{l}}, \ee
 after statistical averaging, decays in time
 \be \label{eq:die}
 \langle\rho\rangle\sim\langle e^{2i\Omega
 t}Q^{-1}[\omega(t)]\rangle=e^{2i\Omega t}\exp\big(-\frac{t}{2}\sqrt{\frac{\pi}{\alpha_{0}}}Erf(t\sqrt{\alpha_{0}})\big)
\ee where $Erf(\ldots)$ is an error function \cite{Handbook}.
    Thus the time of transition from pure to mixed state is
completely determined by the width of autocorrelation function of
random variable $x(t)$: \be\label{eq:auto} \langle
x(t+\tau)x(t)\rangle=e^{-\alpha_{0}\tau^{2}}
 \ee
 In the time interval $0<t<\sqrt{\alpha_{0}/\pi}$, the state of
 the system may be considered as pure. The transition to mixed
 state takes place at $t>\sqrt{\alpha_{0}/\pi}$
 \cite{Chotorlishvili}.
    In order to define level populations of the spin subsystem, we summarize
all possible contributions from the different field states
 (\ref{eq:sol})
 \begin{eqnarray}\label{eq:sybsyst}
 W(t,|11|\rangle)=\sum\limits_{n=0}^{\infty}|C_{1,1,n-1}(t)|^{2},~~~~~~~~~
 W(t,|10|\rangle)=\sum\limits_{n=0}^{\infty}|C_{1,0,n-1}(t)|^{2},\nonumber \\
  W(t,|01|\rangle)=\sum\limits_{n=0}^{\infty}|C_{0,1,n-1}(t)|^{2},~~~~~~~~~~~ W(t,|00|\rangle)=\sum\limits_{n=0}^{\infty}|C_{0,0,n+1}(t)|^{2},
 \end{eqnarray}
    Using Eq.(\ref{eq:sybsyst}), the wave function of spin
 subsystem may be presented in the following form:
\be \label{eq:spinsyb}
|\Psi_{system}\rangle=\sqrt{W(t,|11\rangle)}|11\rangle+\sqrt{W(t,|01\rangle)}|01\rangle+\sqrt{W(t,|10\rangle)}|10\rangle+\sqrt{W(t,|00\rangle)}|00\rangle
\ee
    Not, that quantities $W(t,|i,j\rangle)$
contain terms with random time dependence. But in time interval
$0<t<\sqrt{\alpha_{0}/\pi}$, the system is still in the  pure state
\cite{Chotorlishvili}.
    That is why, for determination of entanglement of spin system
definition for pure states (\ref{eq:state}) should be used.
    Taking into account (\ref{eq:spinsyb}), we obtain:
\be\label{eq:enta}
C(|\Psi(t)\rangle_{S})=2|\sqrt{W(t,|11\rangle)W(t,|00\rangle)}-\sqrt{W(t,|01\rangle)W(t,|10\rangle)}|
\ee

    After lapse of time $t>\sqrt{\alpha_{0}/\pi}$, random
character of quantity $x(t)$ becomes of great importance. Therefore
 (\ref{eq:enta}) should be averaged over all possible realizations of
random variable.

In order to do this, one should replace the quantities
$W(t,|i,j\rangle)$ by their statistically averaged values $\langle
W(t,|i,j\rangle)\rangle$.
    Formation of mixed state in the system, is manifested
in zeroing of time dependent interference terms in $\langle
W(t,|i,j\rangle)\rangle$. Since this means also zeroing of
non-diagonal matrix elements of the density matrix (16). As a
result, statistically averaged concurrence of pure state
$C(|\Psi(t)\rangle)$ will correspond to the concurrence of mixed
state: \be \label{eq:mixed1}
C_{mixed}(|\Psi(t>\sqrt{\alpha_{0}/\pi}))=\langle
C_{pure}(|\Psi\rangle)\rangle\ee
    The expression (\ref{eq:mixed1}) may be considered as
alternative definition of concurrence of mixed states. As a
result, after decay of interference terms
$t>>\sqrt{\alpha_{0}/\pi}$, we obtain: \be \label{eq:mix}
C_{mixed}=2|\sqrt{W_{11}W_{00}}-\sqrt{W_{01}W_{10}}|,\ee
 where
$$W_{ij}=\langle W(t>\sqrt{\alpha_{0}/\pi}|ij\rangle)$$.
    Quantities $W_{ij}$ are not arbitrary constants, by which the
expression (\ref{eq:mix}) needs to be minimized. This is principal
difference of our definition (21) from the standard one (3).
Standard definition has nothing to do with the mechanisms of
formation of mixed state and implies all possible realizations of
it. While, in our case mixed state is exactly defined by the
features of the dynamic. Explicit expressions for $W_{ij}$ were
obtained in \cite{Chotorlishvili}. It turned out that quantities
$W_{ij}$ are expressed in terms of generalized hypergeometric
functions and depend on average number of photons inside of cavity
$W_{ij}(\bar{n})$.

\be
 \langle
 W(t,|1,0\rangle)\rangle=\langle
 W(t,|0,1\rangle)\rangle=\frac{C^{2}}{2}+\frac{C^{2}}{32}\left(12-\frac{2}{\bar{n}}+\frac{\exp(\bar{n})\sqrt{\pi}(1-2\bar{n})^{2}Erfi(\sqrt{\bar{n}})}{\bar{n}^{3/2}}\right),
 \ee

\begin{eqnarray}
\langle W(t,|1,1\rangle)\rangle =
C^{2}\left(\frac{1}{2}-\frac{\exp(-\bar{n}\sqrt{\pi}Erfi(\bar{n}))}{4\bar{n}}\right)+
\frac{C^{2}\exp(-\bar{n})}{2}
\Bigg(\frac{1}{3}(3-2\bar{n}+\bar{n}^{2})\cdot \nonumber \\
\cdot F\left[
\left\{\frac{3}{2},\frac{3}{2}\right\},\left\{\frac{5}{2},\frac{5}{2}\right\},\bar{n}\right]
\Bigg)+\frac{C^{2}\exp(-\bar{n})}{2}\Bigg(\frac{1}{25}(13\bar{n}-2\bar{n}^{2})F\left[
\left\{\frac{5}{2},\frac{5}{2}\right\},\left\{\frac{7}{2},\frac{7}{2}\right\},\bar{n}\right]
\end{eqnarray}
 $$+\frac{3}{49}\bar{n}^{2}F\left[
\left\{\frac{7}{2},\frac{7}{2}\right\},\left\{\frac{9}{2},\frac{9}{2}\right\},\bar{n}\right]\Bigg),$$

$$\langle W(t,|0,0\rangle)\rangle
=C^{2}\left(\frac{1}{2}+\frac{\exp(-\bar{n}\bar{})\sqrt{\pi}Erfi(\bar{n})}{4\bar{n}}\right)+\frac{C^{2}}{1058400\bar{n}^{3/2}}\cdot$$
$$ \cdot \bigg
\{3075\left[2\sqrt{\bar{n}}(-16\bar{n}+\exp(-\bar{n})(3-14\bar{n}-16\bar{n}^{2}))-3\sqrt{\pi}(1-2\bar{n})^{2}Erfi(\sqrt{\bar{n}})\right]+$$
$$ +32 \bar{n}^{3/2}\Bigg[33075\bar{n}F\left[
\left\{\frac{1}{2},\frac{1}{2}\right\},\left\{\frac{3}{2},\frac{3}{2}\right\},\bar{n}\right]+3675(12+5\bar{n}^{2})F\left[
\left\{\frac{3}{2},\frac{3}{2}\right\},\left\{\frac{5}{2},\frac{5}{2}\right\},\bar{n}\right]+$$
$$ +27\bar{n}\bigg[147(14-2\bar{n}+\bar{n}^{2})F\left[
\left\{\frac{5}{2},\frac{5}{2}\right\},\left\{\frac{7}{2},\frac{7}{2}\right\},\bar{n}\right]-50(12+\bar{n})\bar{n}F\left[
\left\{\frac{7}{2},\frac{7}{2}\right\},\left\{\frac{9}{2},\frac{9}{2}\right\},\bar{n}\right]\bigg]+$$
 \be+1225\bar{n}^{3}F\left[
\left\{\frac{9}{2},\frac{9}{2}\right\},\left\{\frac{11}{2},\frac{11}{2}\right\},\bar{n}\right]\Bigg]\Bigg\}.\ee

Explicit form of dependence $W_{ij}(\bar{n})$ gives us possibility
to start from essentially quantum domain $\bar{n}\sim1$ and see
what happens in the semiclassical limit $W_{ij}(\bar{n})$. In
other words, let us proof, how realistic the situation is.

In quasi-classical limit $\bar{n}>>1$, the form of coefficients is
oversimplified \cite{Chotorlishvili}.

 \be
W(t>\sqrt{\alpha_{0}/\pi}|11\rangle)=W(t>\sqrt{\alpha_{0}/\pi}|00\rangle)=
W(t>\sqrt{\alpha_{0}/\pi}|10\rangle)=W(t>\sqrt{\alpha_{0}/\pi}|01\rangle)
 .\ee

As a consequence $C_{mixed}(\bar{n}>>1)=0$. The absence of quantum
correlations in quasi-classical limit proves correctness of
definition (21).

 \end{document}